\documentclass[%
 reprint,
 amsmath,amssymb,
 aps,
]{revtex4-2}
\usepackage[utf8]{inputenc}
\usepackage{graphicx}
\usepackage{dcolumn}
\usepackage{hyperref}
\usepackage{physics}
\usepackage{bm}
\usepackage{hyperref}
\usepackage{subcaption}
\usepackage{xcolor}
\usepackage{comment}
\usepackage{float}
\usepackage{caption}
\begin{document}
\newpage
\textbf{\Large Highlights}

\vspace{0.5cm}

\begin{minipage}{\textwidth}
{\parbox{\linewidth}{
\begin{itemize}
    \item The exact solution to the concentration and segregation of Cr in spherical coordinates is derived.
    \item Segregation in spherical domains exhibits a dose rate dependence that is absent in Cartesian configurations.  
    \item The trend with interface density in Cartesian does not change with dimensionality.
\end{itemize}
}}
\end{minipage}
\clearpage
\preprint{APS/123-QED}

\title{Dependence of Radiation Induced Segregation of Cr on Sink Dimensionality and Morphology in Fe-Cr Alloys}

\author{Mohammadhossein Nahavandian}
\altaffiliation{School of Mechanical and Automotive Engineering, Clemson University.}
\email{mnahava@clemson.edu}
\author{Anter El-Azab}
\altaffiliation{Department of Materials Engineering, Purdue University.}
\author{Enrique Martínez}
\altaffiliation{School of Mechanical and Automotive Engineering, Clemson University.}
\altaffiliation{Department of Materials Science and Engineering, Clemson University.}
\email{enrique@clemson.edu}

\date{\today}

\begin{abstract}
Radiation-induced segregation (RIS) and chemical redistribution in structural alloys can significantly degrade material performance, ultimately leading to failure. In this study, building on previous work by the authors \cite{martinez2018role}, we investigate how the dimensional characteristics of sinks influence solute concentration distributions and segregation behavior. Specifically, we utilize a kinetic Monte Carlo (KMC) model to simulate atomic-scale diffusion and analyze segregation processes in an Fe–3\%Cr alloy. Our analysis includes three representative sink geometries: one-dimensional (1D), two-dimensional (2D), and three-dimensional (3D) planar sinks to capture the effects of sink dimensionality on Cr segregation at grain boundaries (GBs). We also found solutions of concentration and segregation profiles in these cases as well as for a 3D spherical sink. KMC simulations are performed over a range of temperatures to assess thermal effects on Cr redistribution. The results reveal distinct segregation profiles and concentration gradients, although the dependence with sink density seems to remain linear in all cases with planar sinks. The analytical results show that this is not the case in spherical domains, with a more complex dependence of segregation on sink density. Our finite difference solutions for domains including 2D and 3D planer sinks show agreement with corresponding KMC results. 
\\  
\begin{description}
\item[Keywords]
Radiation-Induced Segregation, KMC, Sink Topology and Dimensionality, \\Chemical Redistribution
 
\end{description}
\end{abstract}

\maketitle

\section{\label{sec:intro}Introduction}

Radiation-induced segregation (RIS) is a critical phenomenon affecting the microstructural stability and mechanical properties of Fe-Cr alloys under irradiation, which are widely used in radiation environments \cite{was2007fundamentals, dai2022radiation}. The redistribution of alloying elements under irradiation can lead to the formation of embrittling phases, thereby compromising the material's integrity \cite{martinez2018role}. Understanding the mechanisms driving RIS is essential for the development of radiation-resistant materials.

Recent studies have highlighted the significant influence of sink dimensionality on RIS behavior \cite{DAVID2023}. Sinks, such as dislocations, GBs, and voids, act as sites for defect annihilation and play a pivotal role in the migration and segregation of alloying elements. Hu et al. \cite{Shenyang2021} developed a microstructure-dependent model of radiation-induced segregation (RIS) to investigate the effect of inhomogeneous thermodynamic and kinetic properties of defects on diffusion and accumulation of solute A in AB binary alloys, capable of predicting defect evolution in materials with inhomogeneous thermodynamic and kinetic properties of defects. Chen et al. \cite{CHEN2025120895} irradiated ultra-high purity Fe-Cr with 800 keV protons at a dose rate of \(2 \times 10^5\) dpa/s up to 2 dpa at 3 µm depth at 250, 350, and 450~$^\circ$C. Scanning transmission electron microscopy (STEM) coupled with energy dispersive X-ray spectroscopy (EDS) revealed a transition from heterogeneous Cr segregation around dislocation loops to homogeneous
Cr-rich precipitates in the matrix in alloys with Cr content $>8$ wt.\% at 250~$^\circ$C.
Messina et al.\cite{MESSINA2020166} have stated that for dilute Cr concentrations, global enrichment occurs below 540 K, and depletion above. This temperature threshold grows with solute concentration.

Bouobda Moladje et al. \cite{MOLADJE2022117523} used Phase-Field (PF) modeling for dislocation climb under irradiation and coupled it to point defects and chemical species transport equations. They predicted RIS in Fe-5\%Cr around isolated dislocations or stacking fault configurations like in symmetric tilt GBs. Ma et al. \cite{MA2024119537} investigated the chemical composition changes around extended defects in CoNiCrFe through hybrid Monte Carlo, molecular dynamics (MD), and theoretical studies and found a pronounced Cr enrichment and Co/Ni/Fe depletion around all defect types. In addition, they introduced a correlation between the degree of structural disorder and the chemical segregation/depletion phenomenon in the proximity of extended defects.

Xia et al.\cite{xia2021experimental} performed He ion irradiation at 350~$^\circ$C to study equilibrium segregation and RIS of Cr at GB in reduced-activation ferritic martensitic (RAFM) steels by STEM with an energy-dispersive spectrometer, which matched the calculated results from Mclean with a modified Perk's model. As the temperature increases, the equilibrium Cr segregation decreases monotonically. Patki et al.\cite{PATKI2023154205} studied neutron irradiated FeCrAl alloys to 1.8 dpa at 357~$^\circ$C and found GB Cr enrichment and Al depletion in all alloys that span the 10-13 wt.\% Cr and 5-6 wt.\% Al composition space. Barres et al.\cite{BARRES2022111676} also show W-shaped profiles of Cr concentration across GBs and a heterogeneous precipitation of Cr–C rich particles in GBs planes after irradiation of Fe-13\%Cr. In the study of RIS for Fe-Cr using AKMC method, Senninger et al. \cite{SENNINGER20161} found that the diffusion of vacancies towards sinks leads to Cr depletion and the diffusion of self-interstitials causes Cr enrichment in the vicinity of sinks. For concentrations lower than 15\%Cr, sinks were enriched in Cr for temperatures lower than a threshold of about 700 K. When the temperature is above this threshold, the sinks are depleted in Cr.

Offidani et al. \cite{OFFIDANI2025155533} introduced a model Fe–Ni–Cr alloy, which was irradiated using Fe-ions and quantified the GB chemistry as a function of depth and dose, highlighting that migration on the GB away from the irradiation region can significantly impact solute segregation driven by irradiation.
Sink strengths calculated with object KMC in microstructures containing different types of sinks show that the strengths of vacancies and dislocations span several orders of magnitude over the parameter space, whereas the sink strength of GBs as a function of migration energies of vacancy and interstitial stays constant \cite{JOURDAN2025}.

While the real case reactors are in 3D, the dimensionality of the sink network—whether they are one-dimensional (1D), two-dimensional (2D), or three-dimensional (3D) planar sinks, or spherical sinks, is not fully explored in the literature and the studies mostly focus on the effects of alloying element concentrations, dpa, or temperature on the variation of concentration profiles of solute and whether it is depleting or enriching in the proximity of a sink. In this paper, we investigate the role of sink dimensionality in the radiation-induced segregation of alloying elements in Fe-Cr alloys. Employing simulation techniques and theoretical analysis, we aim to elucidate the underlying mechanisms governing RIS depending on the sink morphology and provide insights into the development of more resilient materials. Our findings reveal that the dimensionality of sinks does not alter the segregation trends for flat boundaries, although their morphology does modify the segregation behavior and its dependence on sink density.

\section{\label{sec:methods}Methods}
\subsection{\label{sec:topology}Effect of topology on segregation}

\begin{figure}[h]
\includegraphics[width=0.45\textwidth]{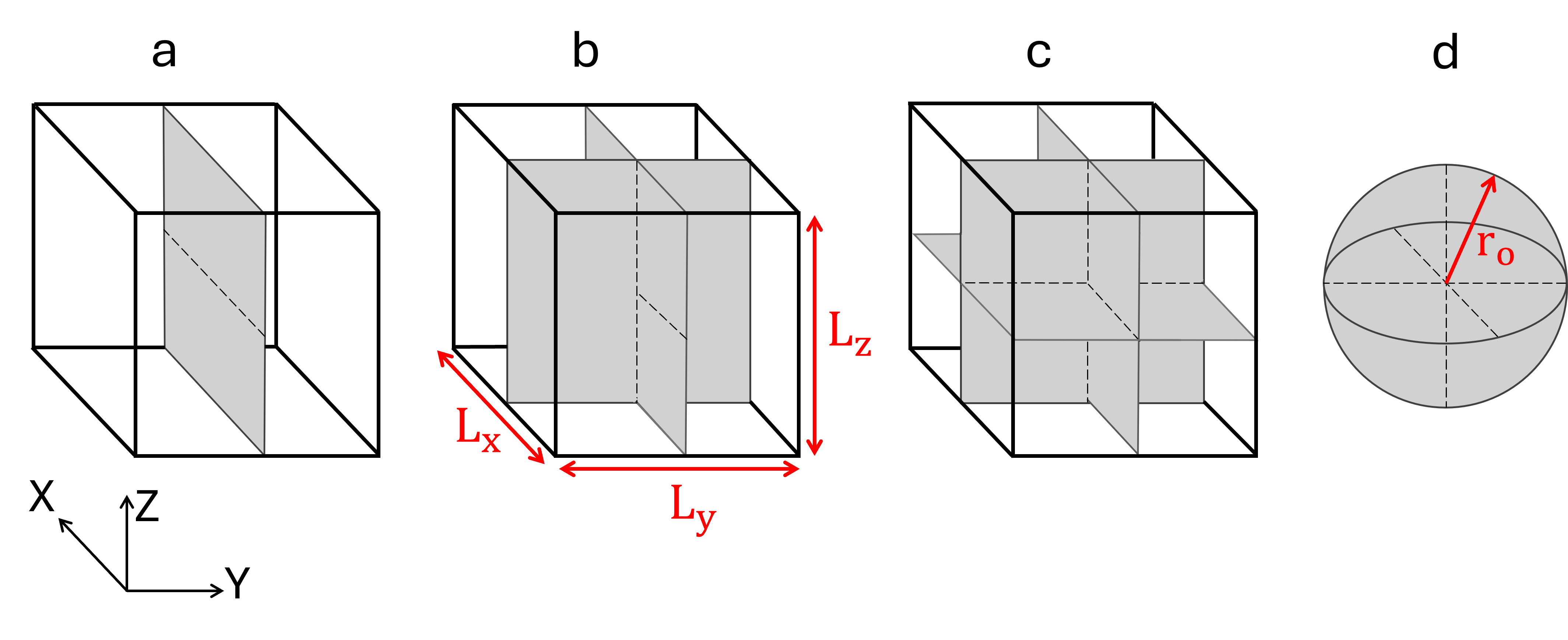}
\caption{\label{fig:schenmatic} Schematic of planar GBs (gray surfaces) acting as sinks in a) 1D, b) 2D c) 3D. d) Spherical domain with the outer boundary acting as an ideal sink.}
\end{figure}

In this section, we study the Cr segregation profile in a spherical domain under irradiation. We obtain a solution of the vacancy concentration in the limit of negligible recombination, i.e., small domains, where the probability of defects being annihilated at sinks is much larger than the recombination probability. Once we have an expression for the vacancy concentration as a function of position, we apply Wiedersich relation \cite{WIEDERSICH1979} to determine the concentration of Cr near sinks, extending the planar interface model from Martinez et al.~\cite{martinez2018role}. The original Wiedersich equation relates the gradient of solute concentration $\nabla c_{\text{Cr}}$ to the vacancy gradient $\nabla c_v$.

\begin{equation}
    \nabla c_{\text{Cr}}(z) = \alpha(z) \frac{\nabla c_v(z)}{c_v(z)},
    \label{wied1}
\end{equation}
where $\alpha(z)=-\frac{L_{Fe,i} L_{Fe,v}}{L_{Fe,i} D_{Cr} + L_{Cr,i} D_{Fe}}(\frac{L_{Cr,v}}{L_{Fe,v}} - \frac{L_{Cr,i}}{L_{Fe,i}})$, with $L_{X,d}$ the transport coefficient of specie $X$ through a $d$ mechanism, and $D_X$ the partial diffusivities of the alloy elements. 
In Figure \ref{fig:schenmatic} we show the different types of sink morphologies used in this study. In 1D, there is only one planar sink with normal in the $x$ direction, in 2D there are two planes perpendicular to each other with normal directions in $x$ and $y$, and in 3D, three planes are perpendicular to each other with normals in $x$, $y$, and $z$ directions (see Figures \ref{fig:schenmatic}-a,b,c). Figure \ref{fig:schenmatic}-d represents the case of a spherical domain with the surface of the domain acting as an ideal sink.

The diffusion equation that we aim to solve for the vacancy concentration is 
\begin{equation}
    \frac{dc_v}{dt} = G + D_v \frac{1}{r^2} \frac{\partial}{\partial r} (r^2 \frac{\partial c_v}{\partial r}) 
\label{eq:1simplified}
\end{equation}
where, due to symmetry, the polar and azimuthal terms cancel. At steady state
\begin{equation}
    \nabla^2 c_v = \frac{1}{r^2} \pdv{r} \left( r^2 \pdv{c_v}{r} \right) = -\frac{G}{D_v},
    \label{eq:laplacian_cv}
\end{equation}
where $G$ is the defect production rate and $D_v$ the vacancy diffusivity. Applying boundary conditions
\begin{align}
    c_v(r_o) &= c_v^{\text{eq}} \quad \text{(at the boundary)}, \\
    \left. \pdv{c_v}{r} \right|_{r \to 0} &= finite \quad \text{(center)},
    \label{eq:BC}
\end{align}
we obtain
\begin{equation}
    c_v(r) = -\frac{G}{6D_v} (r^2 - r_o^2) + c_v^{\text{eq}}.
    \label{eq:spherical-Sol}
\end{equation}

Eq.~\eqref{wied1} in spherical coordinates becomes
\begin{equation}
    \pdv{c_{\text{Cr}}}{r} = \alpha(r) \frac{1}{c_v(r)} \pdv{c_v(r)}{r},
    \label{wied_sp}
\end{equation}
where from Eq.~\eqref{eq:spherical-Sol} we obtain $\pdv{c_v}{r} = -\frac{G}{3D_v} r$. Substituting $c_v(r)$ from Eq.~\eqref{eq:spherical-Sol} in Eq.~\eqref{wied_sp}
\begin{equation}
    \pdv{c_{\text{Cr}}}{r} = \alpha(r) \frac{-\frac{G}{3D_v} r}{-\frac{G}{6D_v} (r^2 - r_o^2) + c_v^{\text{eq}}},
    \label{grad_ccr}
\end{equation}
Integrating Eq.~\eqref{grad_ccr} yields the Cr concentration profile:
\begin{equation}
    c_{\text{Cr}}(r) = \alpha \ln \left( \frac{G}{6D_v} (r_o^2 - r^2) + c_v^{\text{eq}} \right) + \beta.
    \label{eq:c_cr_sph}
\end{equation}
where $\beta$ is a constant that can be found through conservation of mass:
\begin{equation}
\begin{split}
    \int_0^{r_o} \Big[ \alpha \ln \left( \frac{G}{6D_v} (r_o^2 - r^2) + c_v^{\text{eq}} \right) + \beta \Big]4 \pi r^2 dr &\\
    =c_{cr}^{nominal} \frac{4\pi r_o^3}{3}.
\end{split}
\label{eq:mas_conse}
\end{equation}

\begin{figure}[H]
\includegraphics[width=0.45\textwidth]{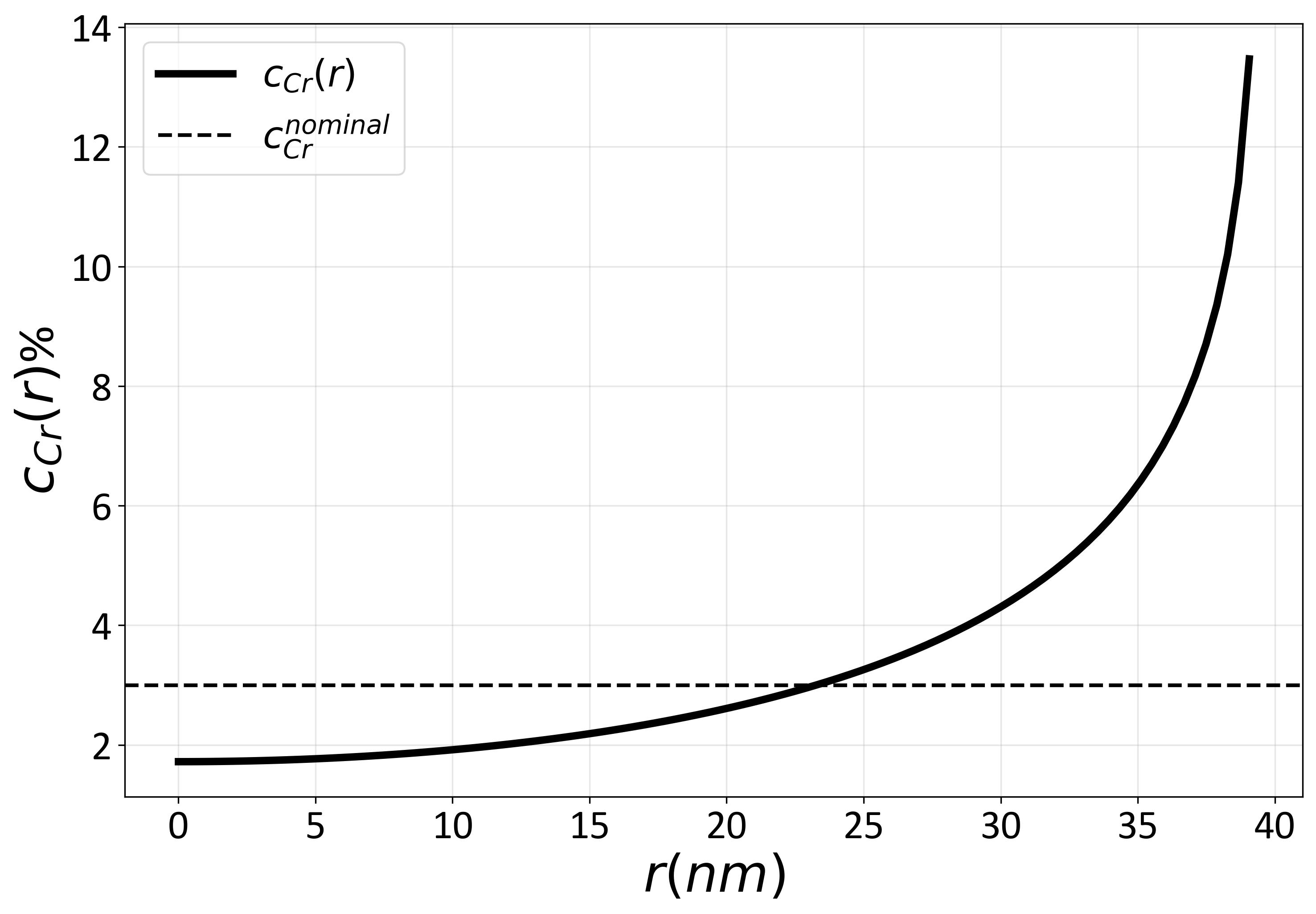}
\caption{\centering\label{fig:C_cr_sp@500K} Steady-state chromium concentration distributions (3 at\% nominal composition) at 500 K, influenced by a spherical sink located at the interface under an irradiation dose rate of $10^{-6}$ dpa/s}
\end{figure}

Assuming $a=\frac{G}{6D_v}$ and $b=c_v^{eq}$, with some algebra, thoroughly explained in the Supplementary Material, we obtain $\beta$ as
\begin{equation}
\begin{split}
     \beta =& c_{cr}^{nominal}
    -\alpha \ln (b) -\frac{ 2\alpha}{r_o^3} \bigg[-\frac{r_o^3}{3} + \frac{(ar_o^2+b)r_o}{a} -\\
    &\frac{(ar_o^2+b)^2}{a^{3/2}}\tanh^{-1} \bigg(r_o \sqrt{a/(ar_o^2+b)}\bigg) \bigg].
\end{split}
\label{eq:K}
\end{equation}
Plugging Eq.~\eqref{eq:K} into Eq.~\eqref{eq:c_cr_sph} we reach the solution for the Cr concentration profile in a spherical domain with absorbing boundary conditions. Figure \eqref{fig:C_cr_sp@500K} shows the Cr concentration variation with sink radius resulting from Eq.\eqref{eq:c_cr_sph_f}.
\begin{equation}
\begin{split}
   &  c_{\text{Cr}}(r) = \alpha \ln \left( \frac{G}{6D_v} (r_o^2 - r^2)
    + c_v^{\text{eq}} \right) 
    + c_{cr}^{nominal} \\
    &-\alpha \ln (c_v^{\text{eq}})
     -\frac{ \alpha}{r_o^3} \bigg[-\frac{r_o^3}{3} + \frac{(\frac{G}{6D_v}r_o^2 +  c_v^{\text{eq}})r_o}{\frac{G}{6D_v}} -\\
    &\frac{(\frac{G}{6D_v} r_o^2 +  c_v^{\text{eq}})^2}{(\frac{G}{6D_v})^{3/2}}\tanh^{-1} \bigg(r_o \sqrt{(\frac{G}{6D_v})\bigg/(\frac{G}{6D_v}r_o^2 +  c_v^{\text{eq}})}\bigg) \bigg].
\end{split}
\label{eq:c_cr_sph_f}
\end{equation}

Figure \eqref{fig:C_cr_sp@500K} shows the change in the Cr concentration at the sink for a spherical system where the maximum radius is equivalent to a cube with $L=63.61$ (nm). 

Assuming planar sinks perpendicular to each other in a cube (shown in Figure \eqref{fig:schenmatic}-b and c), we can follow the same methodology as the spherical system to derive the Cr concentration profiles in Cartesian coordinates. The diffusion equation in a 2D configuration for defects (vacancy and interstitial) in a rectangular domain is
\begin{equation}
D_v \nabla^2 c_v + G = 0 \quad \text{for} \quad (x,y) \in [0, L_x/2] \times [0, L_y/2],
\label{eq:gov}
\end{equation}

\begin{figure}[H]
\includegraphics[width=0.45\textwidth]{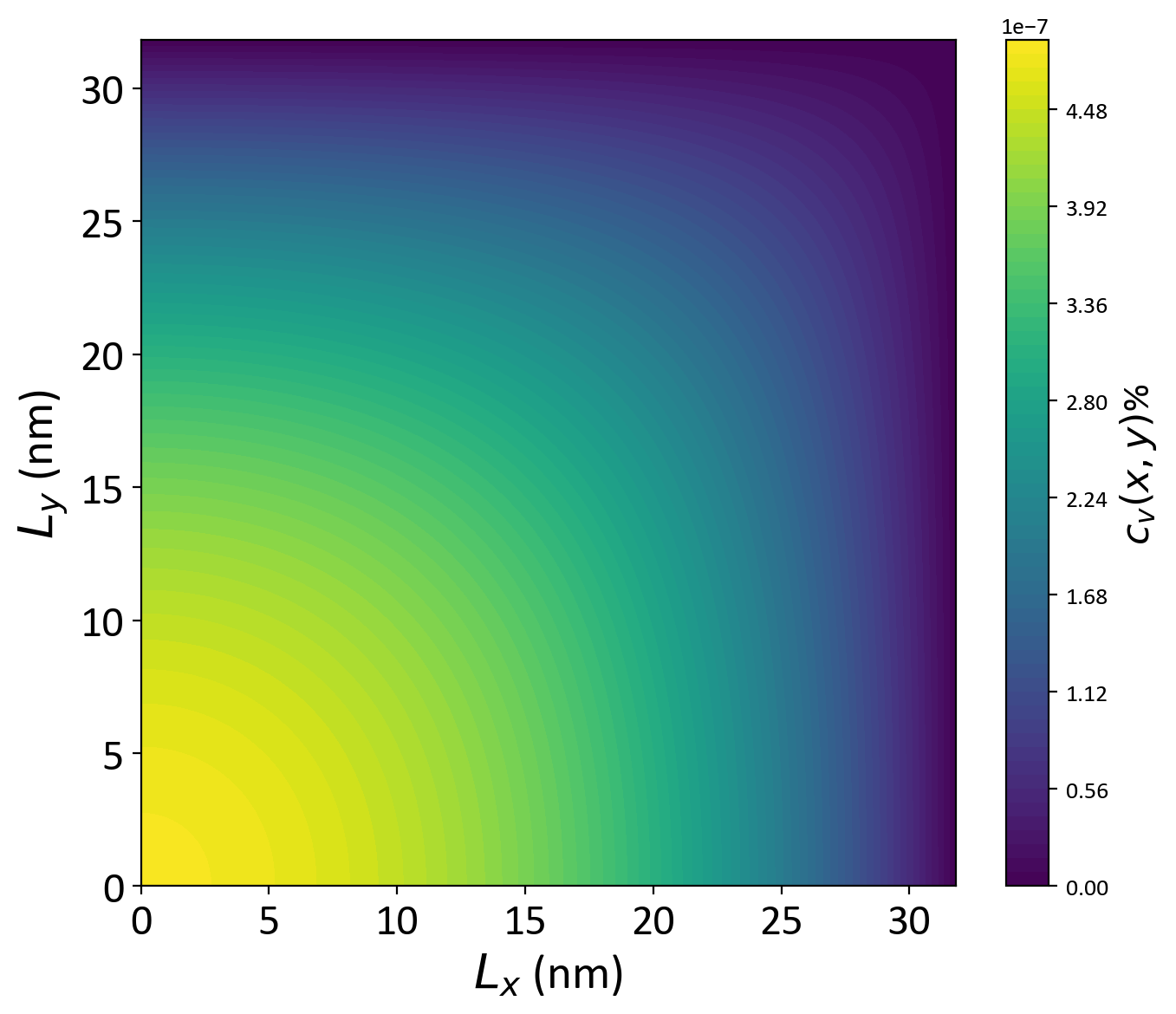}
\caption{\centering\label{fig:Cv_2D@500K} steady-state distribution of vacancy concentration, showing the effect of planar sinks oriented normal to the x and z directions at 500 K under an irradiation dose rate of $10^{-6}$ dpa/s}
\end{figure}

Due to symmetry, we solve for the xy plane in a quarter of the cube. We assume zero flux at $(x,0)$ and $(0,y)$ as well as fixed equilibrium concentration at the sinks located at $x=L_x/2$ and $y=L_y/2$. Applying the Wiedersich formalism, one may obtain the analytical solution for the Cr concentration in the domain, but due to the nature of the problem, a fully analytical solution has proven extremely difficult to obtain, so we used a finite difference (FD) scheme to approximate the solution. Figure \eqref{fig:Cv_2D@500K} shows the vacancy concentration map at 500 K and $G=10^{-6}$ (dpa/s) in a quarter of the domain in the xy plane with a mesh ($100 \times 100$) decaying from its maximum at the origin to the minimum at the intersection of the sink planes. 

\begin{figure}[H]
\includegraphics[width=0.45\textwidth]{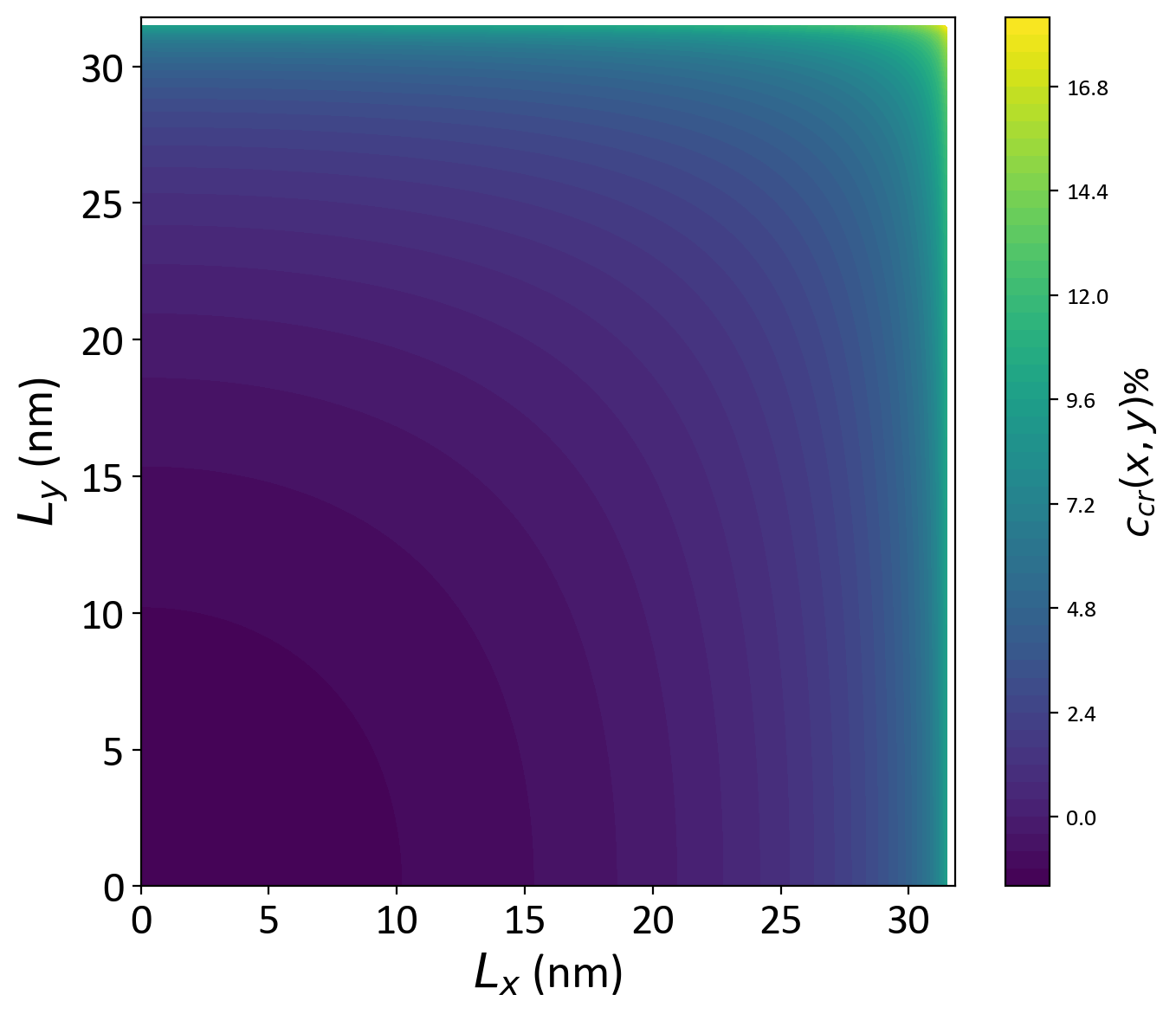}
\caption{\centering\label{fig:C_cr_FD@500K} Steady-state Cr concentration with 3 at\% nominal composition, showing the effect of planar sinks oriented normal to the x and z directions at 500 K under a displacement rate of $10^{-6}$ dpa/s}
\end{figure}

\begin{figure}[H]
\includegraphics[width=0.45\textwidth]{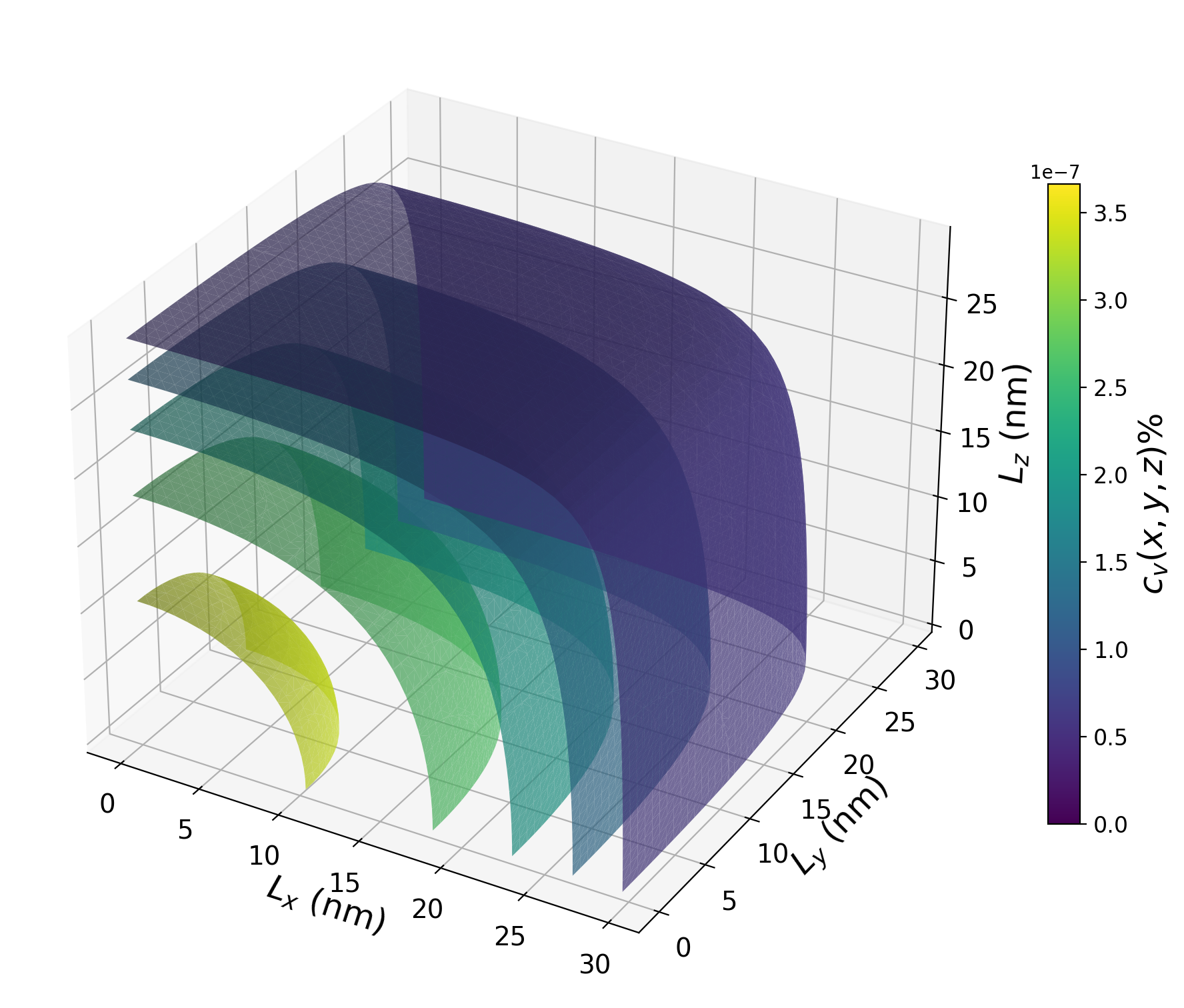}
\caption{\centering\label{fig:C_v_3D_FD@500K} Three-dimensional steady-state distribution of vacancy concentration at 500 K under an irradiation dose rate of $10^{-6}$ dpa/s}
\end{figure}

\begin{figure}[H]
\includegraphics[width=0.45\textwidth]{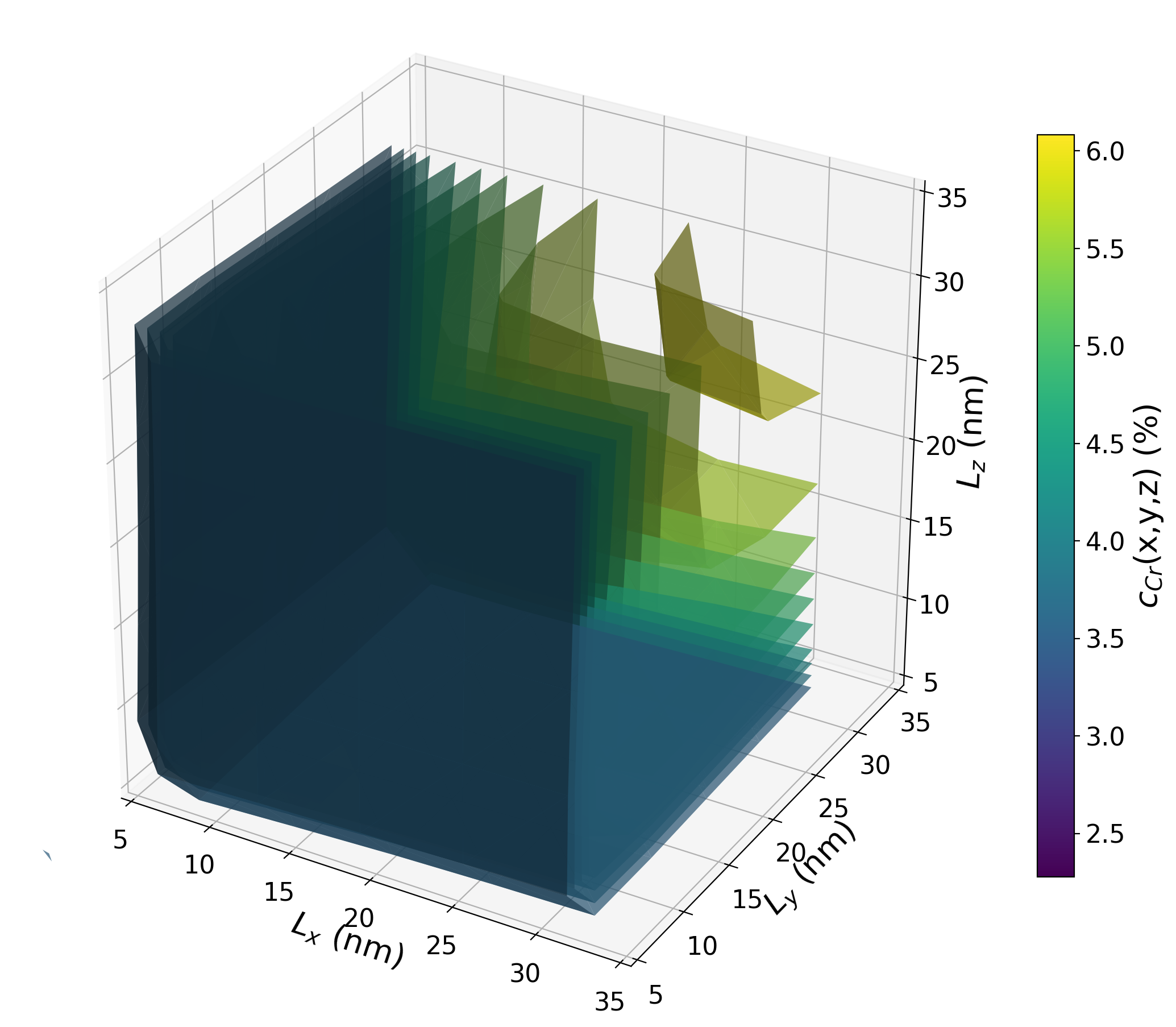}
\caption{\centering\label{fig:C_cr_3D_FD@500K} Three-dimensional steady-state distribution of Cr concentration at 500 K under an irradiation dose rate of $10^{-6}$ dpa/s}
\end{figure}

Figure \eqref{fig:C_cr_FD@500K} is the result of applying the Wiedersich relation to calculate solute concentration and shows the maximum concentration of Cr along the top and right edges located at the sink position with the highest value at the intersection of the sinks. Similarly, for the case of a 3D planar sink morphology, we can find $c_v(x,y,z)$ by solving steady state diffusion in $\quad (x,y,z) \in [0, L_x/2] \times [0, L_y/2] \times [0, L_z/2]$ use the Wiedersich relation to find Cr concentration and its segregation profiles, respectively. Figure \eqref{fig:C_v_3D_FD@500K} shows the vacancy concentrations in a 3D plot, highlighting the decaying trend toward the sink surfaces. In Figure \eqref{fig:C_cr_3D_FD@500K}, the Cr concentration in the cube octants in the presence of three GBs perpendicular to each other is presented. We observe that the maximum concentration is located at the intersection of GBs (triple junction), while the minimum value is observed toward the origin (the farthest from the triple junction), as indicated by the blue iso-surfaces.

\begin{widetext}
\begin{figure}[H]
\centering
\makebox[\textwidth]{
\begin{minipage}{1\textwidth}
\centering
\begin{subfigure}{0.25\textwidth}
\centering
\includegraphics[width=\linewidth]{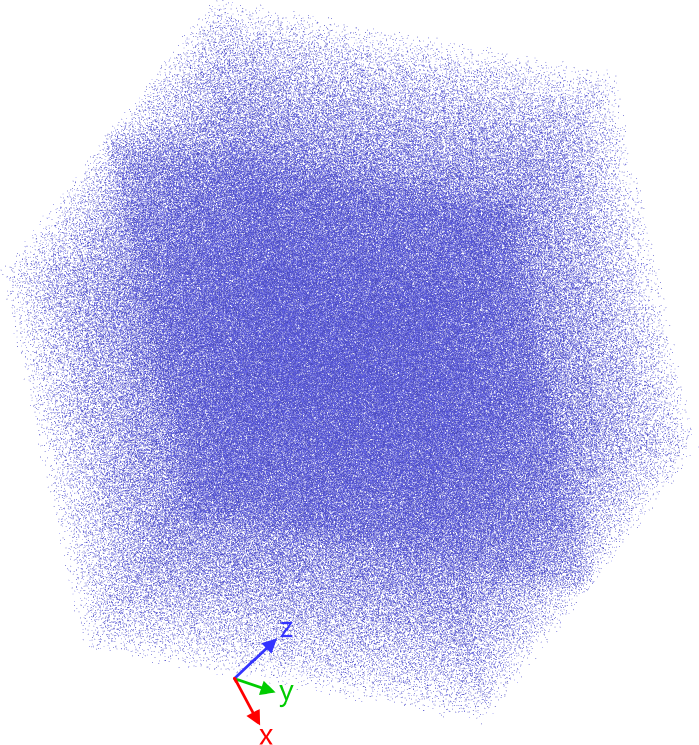}
\caption{One GB normal to z direction}
\label{fig:1D_KMC}
\end{subfigure}
\hfill
\begin{subfigure}{0.25\textwidth}
\centering
\includegraphics[width=\linewidth]{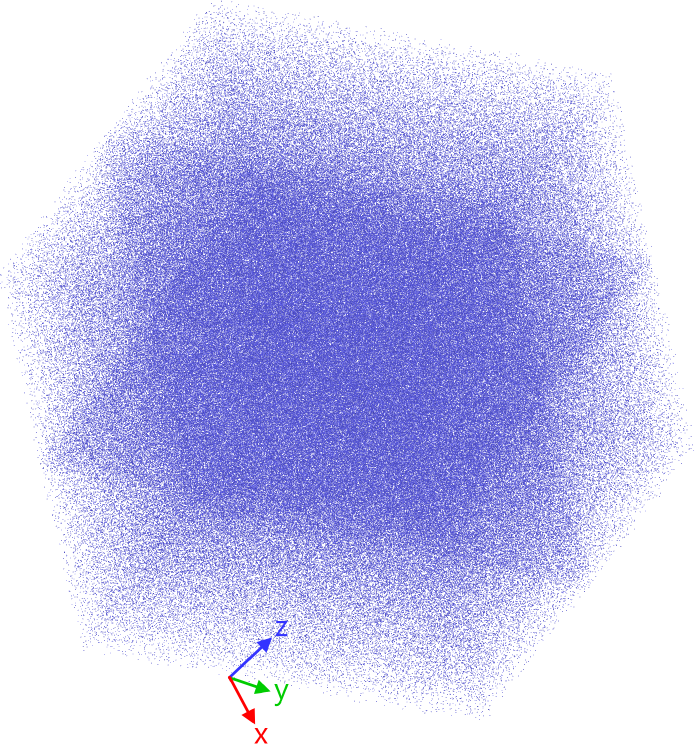}
\caption{Two GBs normal to x and z directions}
\label{fig:2D_KMC}
\end{subfigure}
\hfill
\begin{subfigure}{0.25\textwidth}
\centering
\includegraphics[width=\linewidth]{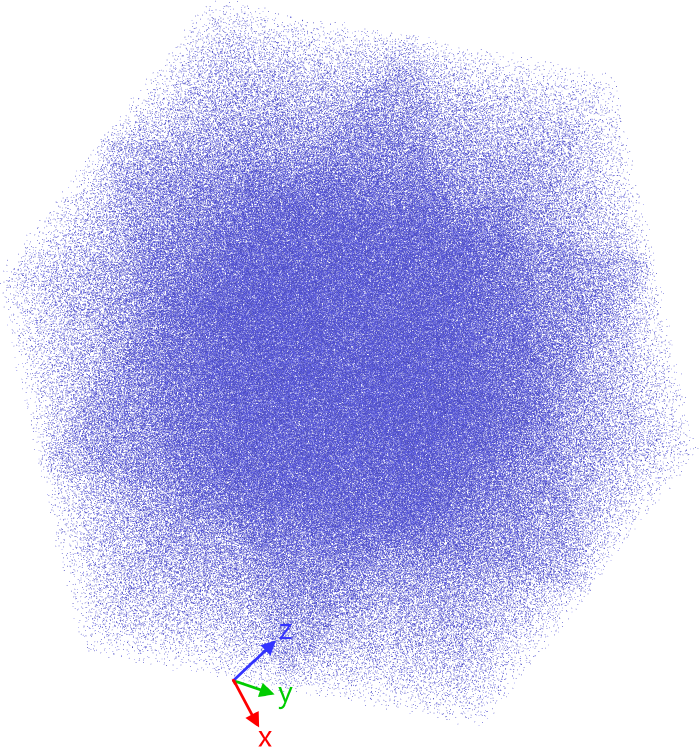}
\caption{Three GBs normal to x, y, and z directions}
\label{fig:3D_KMC}
\end{subfigure}
\caption{Steady-state chromium distribution from KMC simulations at 500 K under $10^{-6}$ dpa/s, showing systems with (a) one, (b) two, and (c) three planar sinks}
\label{fig:KMC_OVITO}
\end{minipage}
}
\end{figure}
\end{widetext}

Figure \eqref{fig:KMC_OVITO} shows snapshots of the Cr distribution at steady state in the KMC simulation at 500 K. Using 3 at\% nominal composition for Cr, we see a higher concentration at the GBs placed in the middle of the domain. A slightly larger concentration at the GB intersections is obtained from the FD but is not apparent from the KMC results.

\begin{figure}[hb]
\includegraphics[width=0.38\textwidth]{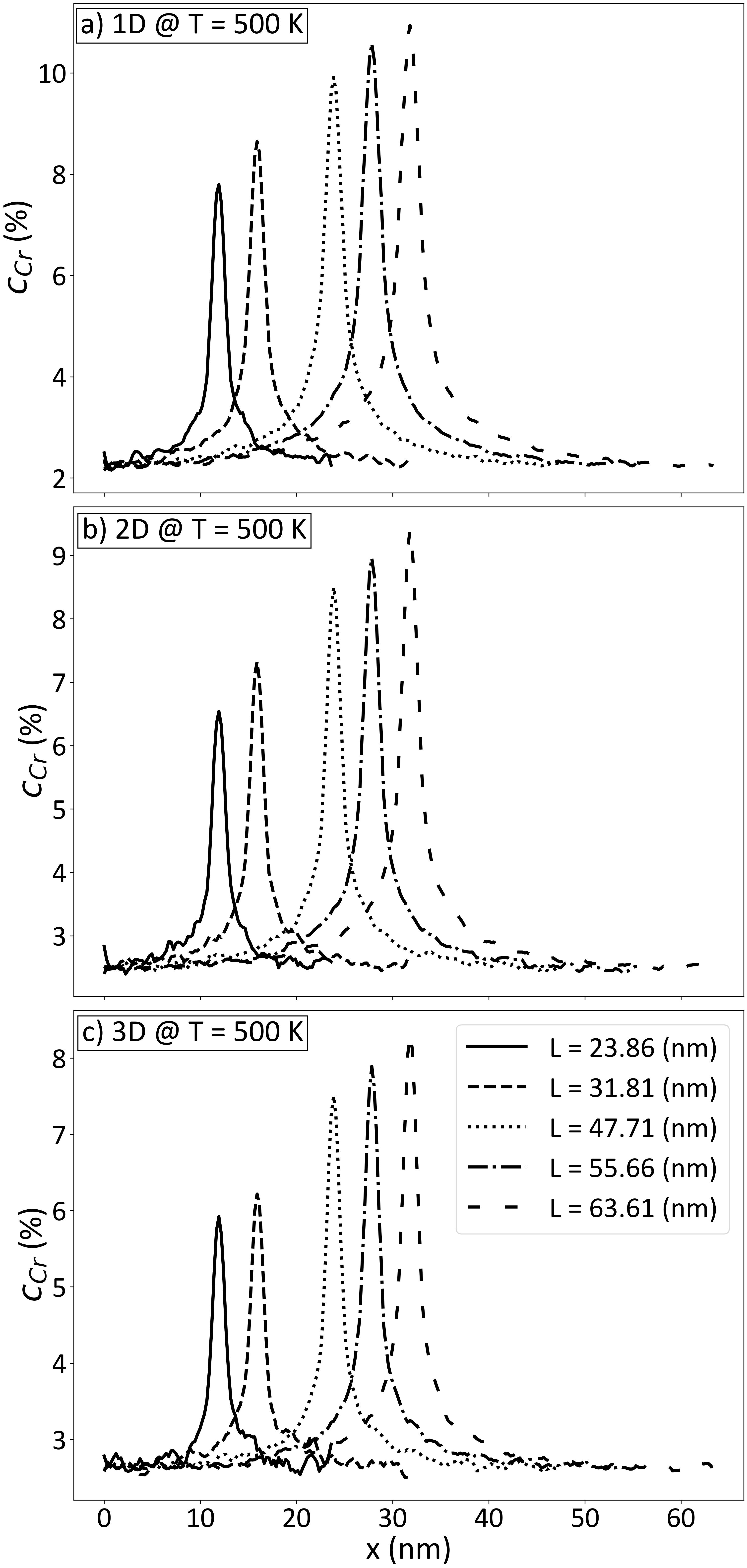}
\caption{\centering\label{fig:C_cr@500K} Steady-state Cr concentration profiles with 0.03  nominal Cr concentration in a) 1D b) 2D, and c) 3D at 500 K}
\end{figure}

\begin{figure}[hb]
\includegraphics[width=0.38\textwidth]{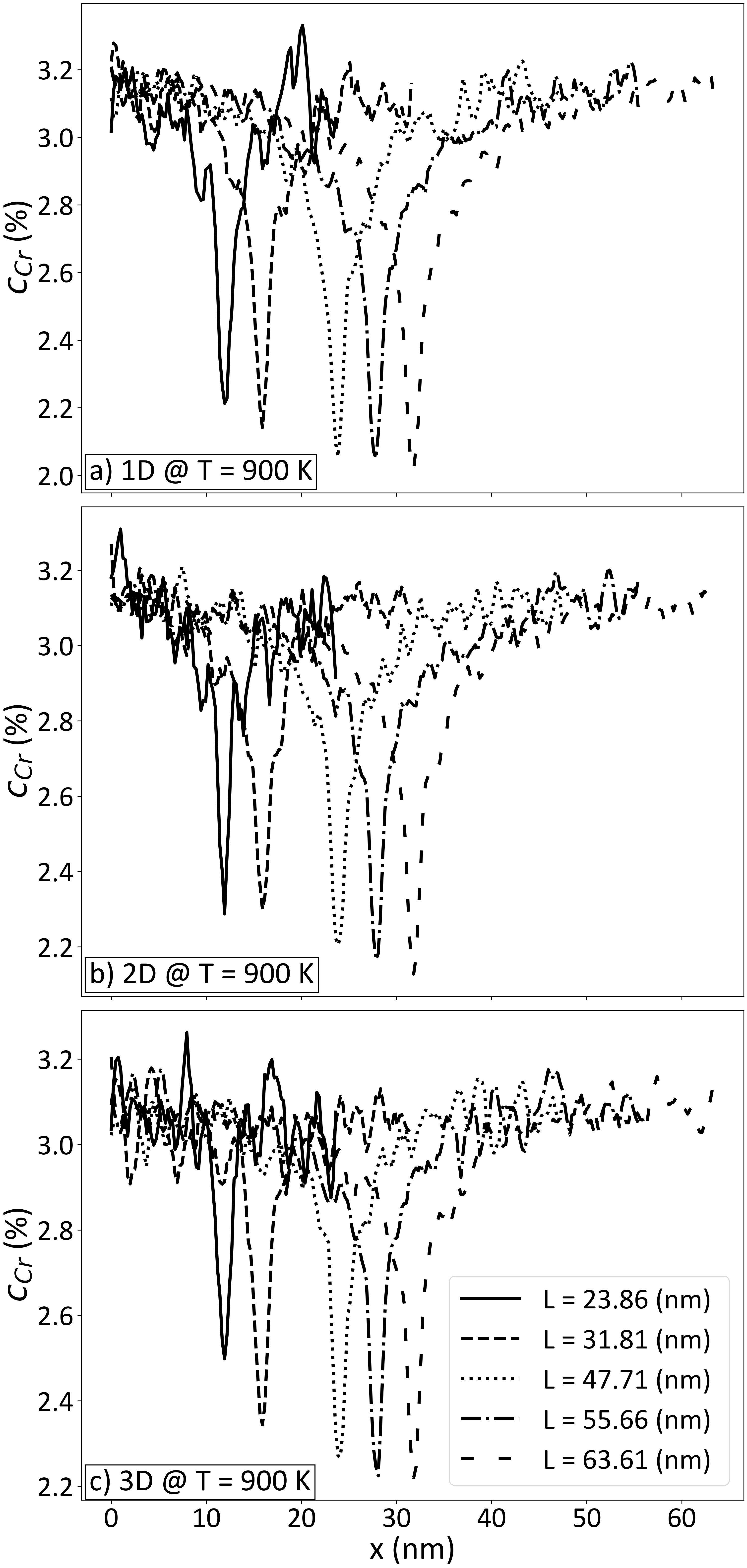}
\caption{\centering\label{fig:C_cr@900K} Steady-state Cr concentration profiles with 0.03  nominal Cr concentration in a) 1D, b) 2D, and c) 3D at 900 K}
\end{figure}

In the KMC computed concentration profiles, shown for relatively low and high temperatures in Figures \eqref{fig:C_cr@500K} and \eqref{fig:C_cr@900K}, we present the average concentration at each z-value. For example, subfigure (a) illustrates that if we have one planar sink with z normal, the concentration values correspond to the average concentration in the xy plane at each z value. By increasing the size of the domain, we see higher maximum concentration magnitude in both cases. We observe an enrichment of Cr at the GB at 500 K, whereas depletion occurs as the temperature increases to 900 K.

The temperature-dependent behavior of Cr under irradiation is attributed to the competition between two distinct point defect migration mechanisms, which dominate at different temperatures. At lower temperatures, the dominant mechanism for Cr segregation is through interstitial migration. 
Irradiation produces excess vacancies and self-interstitials (SIAs). Cr atoms have an attractive binding energy with interstitials, forming mixed dumbbells. These Cr-interstitial pairs migrate more rapidly to sinks, such as GBs.
The preferential flow of Cr via the interstitial flux toward the GB results in an enrichment of Cr in that region. However, at high temperatures, both defects are mobile, but the vacancy mechanism dominates as the propensity for Fe-Cr to stay bound in a SIA decreases.  Vacancies become highly mobile at elevated temperatures.
Due to the inverse Kirkendall effect \cite{WHARRY201442} and the relative size of the atoms, Cr tends to diffuse in the opposite direction of the vacancy flux. As vacancies flow to the GBs, Cr atoms are preferentially swept away from the boundary region. This leads to a depletion of Cr at the GB and a corresponding enrichment of Fe. The crossover from Cr enrichment to depletion occurs at an intermediate temperature where the relative contributions of the interstitial and vacancy mechanisms balance each other which is discussed below.

\section{\label{sec:discussion}Discussion}
In this study, we have employed an analytical approach, FD, and KMC simulations to capture the concentration evolution of Cr in an Fe-3 at\% Cr alloy under irradiation, accounting for different sink dimensionalities and geometries. 
We obtained Cr concentration profiles derived through analytical methods as shown in Figure \eqref{fig:C_cr_sp@500K} and from FD schemes for cartesian systems including 2 and 3 GBs in a 3D domain to find vacancy and Cr concentrations. 
In Figure (\ref{fig:C_cr@500K}), we present Cr concentration profiles for different box sizes with planar ideal sinks in 1D, 2D, and 3D geometries at 500 K, showing Cr segregation. At 900 K (Figure (\ref{fig:C_cr@900K})) Cr depletes the boundary according to the KMC model. 
To compare the KMC and FD results, Figure \eqref{fig:C_cr_FD_KMC@500K} shows average concentration profiles of Cr at 500 K for a sample including 2 planar sinks with $L=63.61 (nm)$. FD values in both cases of vacancy and solute atom concentrations are in agreement with KMC results. Likely due to different defect annihilation propensities induced by time and spatial correlations \cite{MARTINEZ2020} in KMC and FD, the steady-state concentration of vacancies at the interface is an order of magnitude higher than in the KMC with the Cr profile deviating more strongly close to the interface.

\subsection{\label{sec:seg_sp}Segregation in Spherical Coordinates}
In this section, we derive the total Cr segregation expression in a spherical system near the sink. The total amount of segregation is defined by the integral in Eq. \ref{eq:segregation}
\begin{equation}
\begin{aligned}
S^{Cr}_{sp} &= \int_{V} \big[c_{Cr}(r) - c_{Cr}(0) \big] dV \\
&
= 4\pi\int_{0}^{r_o} \big[c_{Cr}(r) - c_{Cr}(0) \big]  r^2 dr \\
&
=4\pi\alpha \int_{0}^{r_o} \bigg[ \ln \left( \frac{G}{6D_v} (r_o^2 - r^2) + c_v^{\text{eq}} 
\right)\\
&-  \ln \left( \frac{G}{6D_v} (r_o^2) + c_v^{\text{eq}} 
\right) \bigg] r^2 dr
\end{aligned}
\label{eq:segregation}
\end{equation}
As explained in the Supplementary Material, we can write the final expression as:

\begin{equation}
\begin{split}
S^{Cr}_{sp}
&=4\pi\alpha\Bigg[
\frac{r_o^3}{3}\ln\!\Big(\frac{c_v^{\mathrm{eq}}}{\dfrac{G}{6D_v} r_o^2 + c_v^{\mathrm{eq}}}\Big)\\
&- \frac{2r_o^3}{9}
- \frac{2r_o (\dfrac{G}{6D_v} r_o^2 + c_v^{\mathrm{eq}})}{3\frac{G}{6D_v}}\\
&+ \frac{2(\dfrac{G}{6D_v} r_o^2 + c_v^{\mathrm{eq}})^{3/2}}{3(\dfrac{G}{6D_v})^{3/2}}\,\operatorname{atanh}\!\Big(r_o\sqrt{\frac{\dfrac{G}{6D_v}}{\dfrac{G}{6D_v} r_o^2 + c_v^{\mathrm{eq}}}}\Big)
\Bigg].
\end{split}
\label{eq:segregation3_simplified}
\end{equation}

If we write $r_o$ as a function of the interface density $\rho_I=A/V$:
\begin{equation}
    r_o = \frac{3}{ \rho_I}
    \label{eq:interface_density}
\end{equation}
Eq.\eqref{eq:segregation3_simplified} can be written as:

\begin{equation}
\begin{split}
S^{Cr}_{sp} &= 4\pi\alpha \Bigg[ 
 \frac{9}{\rho_I^3} \ln\left( \frac{c_v^{\mathrm{eq}}}{\dfrac{9G}{6D_v \rho_I^2} + c_v^{\mathrm{eq}}} \right) 
- \frac{24}{\rho_I^3} 
- \frac{12 D_v c_v^{\mathrm{eq}}}{G \rho_I} \\
& + \frac{2 \left( \dfrac{9G}{6D_v \rho_I^2} + c_v^{\mathrm{eq}} \right)^{3/2}}{3 \left( \dfrac{G}{6D_v} \right)^{3/2}} 
\operatorname{atanh}\left( \frac{3}{\rho_I} \sqrt{ \frac{ \dfrac{G}{6D_v} }{ \dfrac{9G}{6D_v \rho_I^2} + c_v^{\mathrm{eq}}} } \right)
\Bigg].
\end{split}
\label{eq:segregation_interface_density}
\end{equation}

\noindent where the relation with the interface density is more complex than the inversely proportional dependence obtained for the Cartesian morphologies described below. Another important point is that in this last case with spherical domains, the segregation seems to depend on the dose rate, contrary to the Cartesian cases.

\subsection{Segregation in Cartesian Coordinates}
It was shown in Ref \cite{martinez2018role} that the segregation in 1D can be written as 

\begin{equation}
\lim _{c_v^{eq} \rightarrow 0} S_{1D} = \frac{\alpha}{\rho _{Ix}}\left ( \ln 2 -1 \right )
\end{equation}
with $\rho _{Ix}$ the density of planar ideal sinks in 1D.
Here, we calculate the profiles for the cases with 2D and 3D planar sinks. In 2D, the total segregation is given by:

\begin{equation}
S_{2D} = \int_0^{L_y/2} \int_0^{L_x/2} \left[ \ c_{Cr}(x, y) \ - \ c_{Cr}(0, 0) \ \right]  dx  dy,
\label{eq:s2d}
\end{equation} 
We define \(S_{2D}\) as the integrated deviation of the local concentration \(c_{Cr}(x,y)\) from the bulk concentration \(c_{Cr}(0,0)\) over $L_xL_y/4$. The computed value is independent of z. Following the same methodology, the segregation profile in the presence of planar ideal sinks in 3D is given by
\begin{equation}
\begin{aligned}
    S_{3D} &= \int_0^{L_z/2} \int_0^{L_y/2} \int_0^{L_x/2} \Big[ c_{Cr}(x, y, z) \\
    &\quad - c_{Cr}(0,0,0) \Big]  dx  dy dz,
    \label{eq:s3d}    
\end{aligned}    
\end{equation}
where we define \(S_{3D}\) as the integrated deviation of the local concentration \(c_{Cr}(x,y,z)\) from the bulk concentration \(c_{Cr}(0,0,0)\) across $1/8$ of the volume \(V_{total}\). We can also write $\rho _{I_x}=\frac{1}{L_x}$, $\rho _{I_y}=\frac{1}{L_y}$, and $\rho _{I_z}=\frac{1}{L_z}$ as the interface densities in each dimension, with the total density been the sum of these 3 components. Similarly to what has been explained in section (\ref{sec:topology}), we have applied the FD scheme to compute the segregation of Cr toward the planar sinks in these 2D and 3D structures. 

\begin{figure}[hb]
\includegraphics[width=0.45\textwidth]{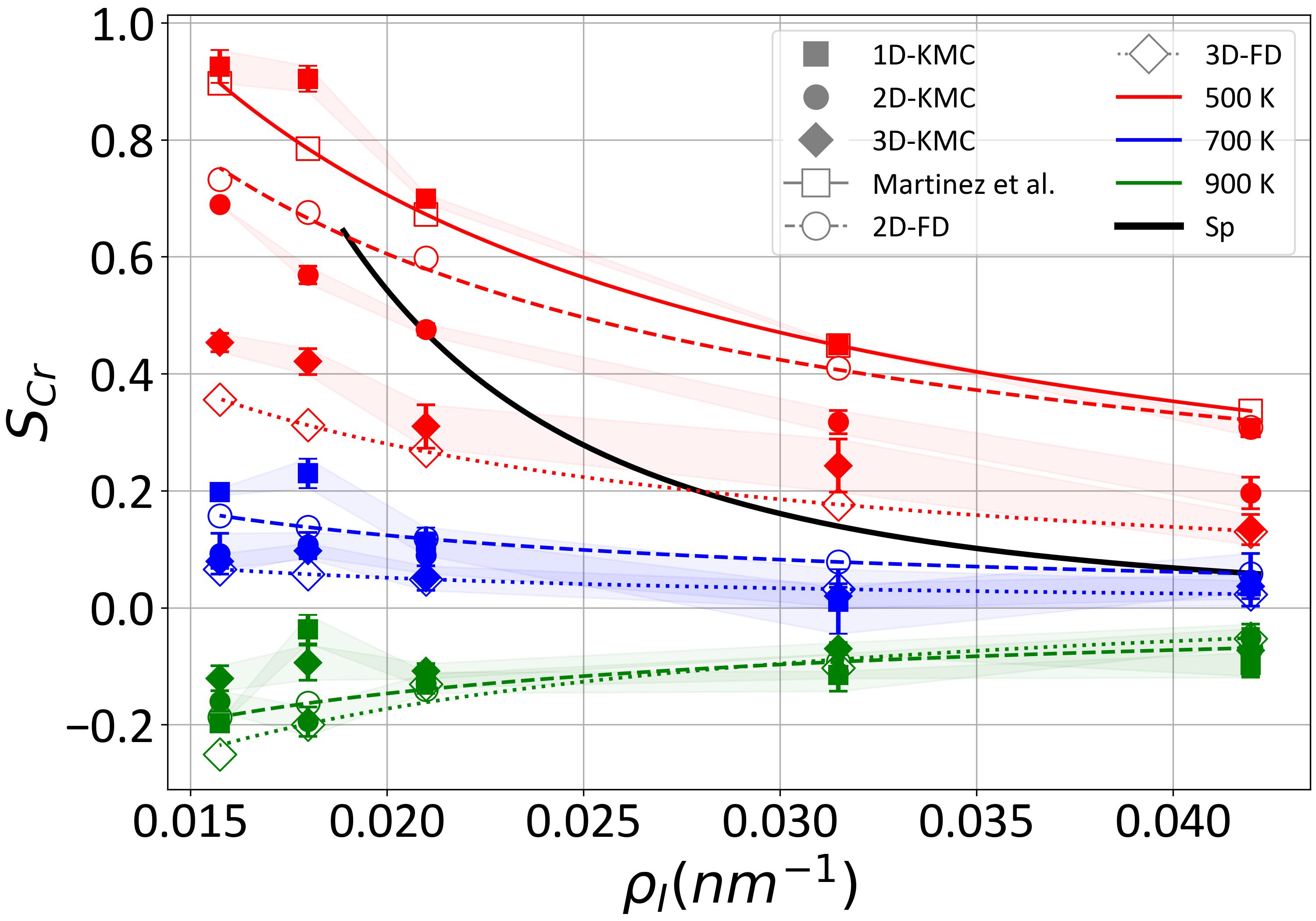}
\caption{\centering\label{fig:Segregation_all} Total steady-state Cr segregation with 0.03  nominal Cr concentration for 1D, 2D and 3D planar ideal sink morphologies at 500, 700 and 900 K computed by KMC, FD, and analytical expression for spherical sink vs interface density.}
\end{figure}

In Figure (\ref{fig:Segregation_all}), the total segregation versus the interface density for 1D, 2D, and 3D morphologies at different temperatures (500, 700, and 900 K) is shown, computed with FD, KMC, and analytical methods.  For the cases of 2D and 3D segregations from FD, we compute the average concentration of Cr at each plane with the normal in the z direction and compare it with the KMC. 
 
This is the reason why we observe a decrease in segregation in 2D with respect to 1D and in 3D with respect to 1D and 2D. The filled scattered points represent KMC data based on 75 simulations with 5 realizations at each temperature and box size. The hollow points are FD-predicted data, which show good agreement with KMC in all conditions. In the FD cases, we have fitted $\alpha$ as -0.02, -0.011 and 0.013 at 500 K, 700 K, and 900 K, respectively. The black solid curve is the segregation in a spherical system, which clearly shows deviation from a linear trend, highlighting that the sink topology can alter the segregation behavior. 

For all dimensionalities, there is clear Cr segregation toward the sink at 500 K. This effect becomes weaker at 700 K and turns into Cr depletion at 900 K. Curves fitted to the KMC data show a clear trend in all cases, regardless of the dimensionality effect. Compared to Figure 2 in Ref \cite{DAVID2023}, we do not see any nonlinear relation between segregation and interface spacing in the case of planar sinks. It is worth reminding that these results assume a negligible defect recombination and that defect annihilation is dominated by their interactions with planar sinks.
To investigate this in more detail, we present a case study of 2D concentration profiles of vacancies and Cr in both FD and KMC, where we observe agreement between the data points. Adding the effect of the recombination of vacancies and SIAs, even with strong recombination rates, did not affect the FD segregation results, and it is more sensitive to the $\alpha$ selection.

\begin{widetext}
\centering
\makebox[\textwidth]{
\begin{minipage}{0.76\textwidth}
\centering
\includegraphics[width=\linewidth]{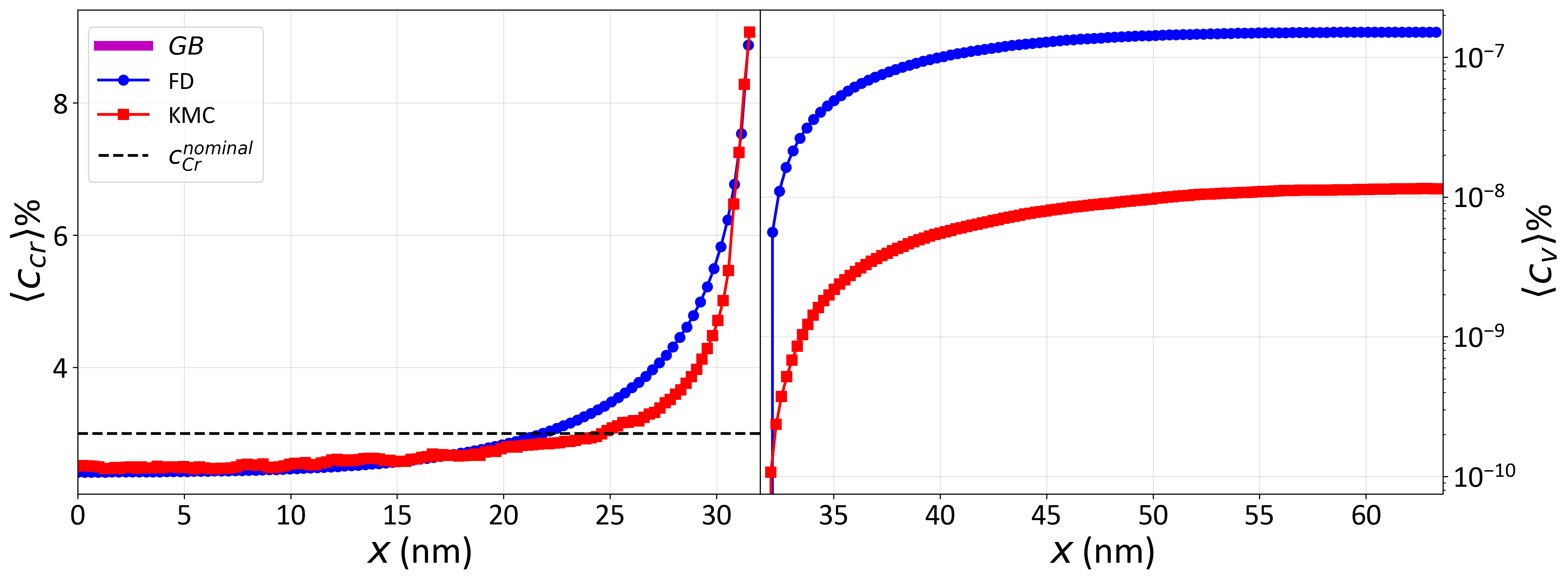}
\captionof{figure}{\label{fig:C_cr_FD_KMC@500K} Steady-state Cr concentration comparison between FD and KMC with 0.03 nominal Cr concentration in 2D at 500 K}
\end{minipage}
}
\end{widetext}

\section{\label{sec:conclusion}Conclusion}
In this paper, we investigated the RIS behavior of Fe-Cr alloys, accounting for different sink dimensionalities, morphologies, and interface spacings across various temperature ranges with a finite differences (FD) scheme, KMC simulations, and analytical approaches. Although the dimension of the sink affects the total amount of Cr segregation, the increasing trend of the segregation with interface spacing indicates a linear behavior in each independent dimension. Spherical domains show a more complex relation between sink density and total segregation. Furthermore, segregation in spherical domains exhibits a dose rate dependence that is absent in Cartesian configurations.  

\section{\label{sec:dec}Declaration of competing interest}
The authors declare that they have no known competing financial interests or personal relationships that could have appeared to influence the work reported in this paper.

\section{\label{sec:ackn}Acknowledgment}

The authors acknowledge the support from the US Department of Energy, Office of Science, Office of Fusion Energy Sciences under the grant number \textbf{DE-SC0024515}. Additionally, this material is based on work supported by the National Science Foundation under Grant Nos. MRI\# 2024205, MRI\# 1725573, and CRI\# 2010270 for allotment of compute time on the Clemson University Palmetto Cluster.

\clearpage
\nocite{*}

\bibliography{ref}
\end{document}


\title{Supplementary Material for: \\ 
\textbf{Dependence of Radiation Induced Segregation of Cr on Sink Dimensionality and Morphology in Fe-Cr Alloys}}
\author{Mohammadhossein Nahavandian$^{1}$, Anter El-Azab$^{2}$, Enrique Martinez$^{1,3}$ \\
\small $^{1}$School of Mechanical and Automotive Engineering, Clemson University \\
\small $^{2}$Department of Materials Science and Engineering, Purdue University \\
\small $^{3}$Department of Materials Science and Engineering, Clemson University \\
\small \texttt{enrique@clemson.edu,mnahava@clemson.edu}\\
\vspace{20mm}
\today}
\maketitle
\clearpage
\section{Cr concentration profile in Spherical coordinates}
\label{sph_sink_sec}
\begin{figure}[h]
\centering
\includegraphics[width=0.9\textwidth]{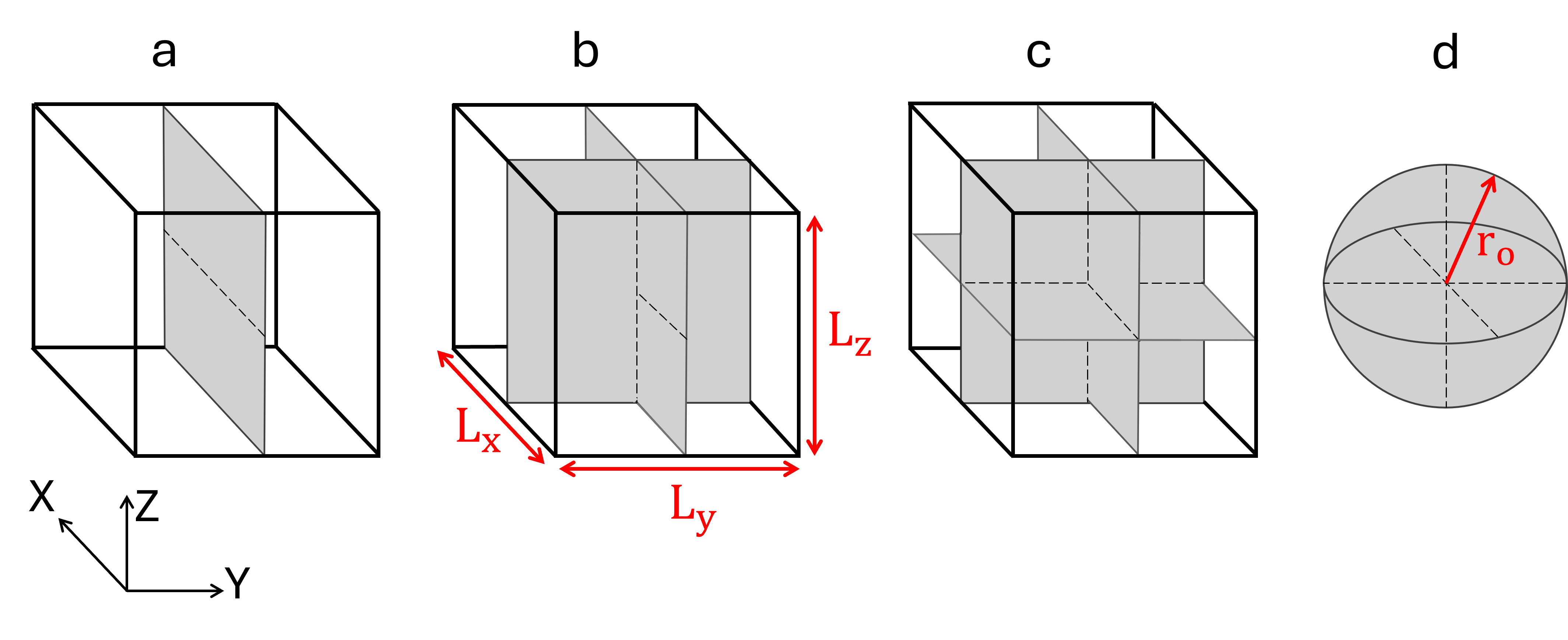}
\caption{Schematic of planar GBs (gray surfaces) acting as sinks in a) 1D, b) 2D c) 3D. d) Spherical domain with the outer boundary acting as an ideal sink.}
\label{fig:schem}
\end{figure}

In this part, we derive the full derivation of the Cr concentration profile in spherical coordinates, starting with the Wiedersich relation \cite{martinez2018role,WIEDERSICH1979} for RIS near spherical sinks. The original equation relates the gradient of solute concentration $\nabla c_{\text{Cr}}$ to the vacancy gradient $\nabla c_v$

\begin{equation}
    \nabla c_{\text{Cr}} = \alpha \frac{\nabla c_v}{c_v},
    \label{eq:Wiedersich}
\end{equation}
where $\alpha=-\frac{L_{\text{Fe},i} L_{\text{Fe},v}}{L_{\text{Fe},i} D_{\text{Cr}} + L_{\text{Cr},i} D_{\text{Fe}}}\left(\frac{L_{\text{Cr},v}}{L_{\text{Fe},v}} - \frac{L_{\text{Cr},i}}{L_{\text{Fe},i}}\right)$. The vacancy concentration $c_v$ around a spherical sink of radius $r$ solves $\nabla^{2} c_{v}$ as (Eq.~\eqref{eq:laplacian}):

\begin{equation}
\begin{split}
    \nabla^2 c_v = &\frac{1}{r^2} \frac{\partial}{\partial r} \left(r^2 \frac{\partial c_v}{\partial r}\right) + 
    \frac{1}{r^2 \sin\theta} \frac{\partial}{\partial \theta} \left(\sin\theta \frac{\partial c_v}{\partial \theta}\right) + \frac{1}{r^2 \sin^2 \theta} \frac{\partial^2 c_v}{\partial \phi^2}, 
\end{split}
\label{eq:laplacian}
\end{equation}
For a system with radial symmetry, we express gradients in spherical coordinates ($r$, $\theta$, $\phi$). Assuming angular independence ($\partial/\partial\theta = \partial/\partial\phi = 0$), the gradient of a scalar field $f(r)$ reduces to:

\begin{equation}
    \nabla f \rightarrow \frac{\partial f}{\partial r} \hat{r},
    \label{eq:gradient}
\end{equation}
assuming conditions mentioned, we can write the unsteady diffusion equation of vacancy as:

\begin{equation}
    \frac{dc_v}{dt} = G + D_v \frac{1}{r^2} \frac{\partial}{\partial r} \left(r^2 \frac{\partial c_v}{\partial r}\right), 
\label{eq:1simplified}
\end{equation}
where $G$ is the defect production rate and $D_v$ the vacancy diffusivity. Figure (\ref{fig:schem}-d) shows the sink geometry we have studied in section \ref{sph_sink_sec}. 

and for steady state:

\begin{equation}
    \nabla^2 c_v = \frac{1}{r^2} \frac{\partial}{\partial r} \left( r^2 \frac{\partial c_v}{\partial r} \right) = -\frac{G}{D_v},
    \label{eq:laplacian_cv}
\end{equation}
With boundary conditions:

\begin{align}
    c_v(r_o) &= c_v^{\text{eq}} \quad \text{(at boundary)}, \\
    \left. \frac{\partial c_v}{\partial r} \right|_{r \to 0} &= \text{finite} \quad \text{(center)},
     \label{eq:BC}
\end{align}

the solution is:

\begin{equation}
    c_v(r) = -\frac{G}{6D_v} (r^2 - r_o^2) + c_v^{\text{eq}},
    \label{eq:spherical-Sol}
\end{equation}

The Wiedersich relation in spherical coordinates becomes:

\begin{equation}
    \frac{\partial c_{\text{Cr}}}{\partial r} = \alpha(r) \frac{1}{c_v(r)} \frac{\partial c_v}{\partial r},
    \label{eq:wieder-sph}
\end{equation}

where $\frac{\partial c_v}{\partial r} = -\frac{G}{3D_v} r$. Substituting $c_v(r)$ and $\frac{\partial c_v}{\partial r}$ in Eq.~\eqref{eq:wieder-sph}:

\begin{equation}
    \frac{\partial c_{\text{Cr}}}{\partial r} = \alpha(r) \frac{-\frac{G}{3D_v} r}{-\frac{G}{6D_v} (r^2 - r_o^2) + c_v^{\text{eq}}},
\end{equation}

Assuming $\alpha$ constant and integrating yields the Cr profile:

\begin{equation}
    c_{\text{Cr}}(r) = \alpha \ln \left( \frac{G}{6D_v} (r_o^2 - r^2) + c_v^{\text{eq}} \right) + \beta,
    \label{eq:c_cr_sph}
\end{equation}

where $\beta$ is a constant that can be found through mass conservation:

\begin{equation}
\begin{split}
    \int_0^{r_o} \left[ \alpha \ln \left( \frac{G}{6D_v} (r_o^2 - r^2) + c_v^{\text{eq}} \right) + \beta \right]4 \pi r^2 dr =c_{\text{Cr}}^{\text{nominal}} \frac{4\pi r_o^3}{3},
\end{split}
\label{eq:mas_conse}
\end{equation}

Where the LHS in Eq.~\eqref{eq:mas_conse} is:

\begin{equation}
\begin{split}
    \alpha \int_0^{r_o} \ln \left( \frac{G}{6D_v} (r_o^2 - r^2) + c_v^{\text{eq}} \right) 4\pi r^2 dr + \frac{4 \beta \pi r_o^3}{3},
\end{split}
\label{eq:LHS}
\end{equation}

we define $a = \frac{G}{6D_v}$ and $b = c_v^{\text{eq}}$:

\begin{equation}
\begin{split}
    \alpha 4\pi \int_0^{r_o} \ln \left( a (r_o^2 - r^2) + b \right) r^2 dr + \frac{4 \beta \pi r_o^3}{3},
\end{split}
\label{eq:LHS1}
\end{equation}

Let's name the first term in Eq.~\eqref{eq:LHS1} as $I=\int_0^{r_o} \ln \left( a (r_o^2 - r^2) + b \right)r^2 dr$. If we change variables to $m=r^3/3$, then $dm=r^2dr$. We can use integration by parts where $n=\ln (a (r_o^2 - r^2) + b)$, $dn=\frac{-2ar}{a (r_o^2 - r^2) + b}dr$. 

\begin{equation}
\begin{split}
    I=&\frac{r^3}{3} \ln \left( a (r_o^2 - r^2) + b \right) \Bigg|_0^{r_o}
    - \int_0^{r_o} \frac{r^3}{3} \frac{-2ar}{a (r_o^2 - r^2) + b}dr\\
    &
    =\frac{r_o^3}{3} \ln (b) + \frac{2a}{3}\int_0^{r_o} \frac{r^4}{a (r_o^2 - r^2) + b}dr,
\end{split}
\label{eq:LHS_I}
\end{equation}

Now, let's define $q=\frac{r_o^3 \ln (b)}{3}$, $c=ar_o^2 +b$:

\begin{equation}
\begin{split}
    I=& q + \frac{2a}{3}\int_0^{r_o} \frac{r^4}{c - ar^2}dr\\
    &
    = q  + \frac{2a}{3}\int_0^{r_o} \left(\frac{-r^2}{a} + \frac{c}{a^2} -\frac{c^2}{a^2(c - ar^2)}\right) dr\\
    &
    =q + \frac{2a}{3} \left(\frac{-r_o^3}{3a} + \frac{cr_o}{a^2} - \frac{c^2}{a^{5/2}} \tanh^{-1} \left(r_o \sqrt{a/c}\right)\right),
\end{split}
\label{eq:LHS_II}
\end{equation}

finally:

\begin{equation}
\begin{split}
    \text{LHS}=& \frac{4\pi \alpha r_o^3 \ln (b)}{3} + \frac{8\pi \alpha}{3} \Bigg[-\frac{r_o^3}{3} + \frac{(ar_o^2+b)r_o}{a} -\\
    &\frac{(ar_o^2+b)^2}{a^{3/2}}\tanh^{-1} \left(r_o \sqrt{a/(ar_o^2+b)}\right) \Bigg]
    + \frac{4\pi \beta r_o^3}{3} = c_{\text{Cr}}^{\text{nominal}} \frac{4\pi r_o^3}{3},
\end{split}
\label{eq:LHS_III}
\end{equation}

so we can find the constant $\beta$ from Eq.~\eqref{eq:LHS_III}:

\begin{equation}
\begin{split}
     \beta =& c_{\text{Cr}}^{\text{nominal}}
    -\alpha \ln (b) -\frac{ 2\alpha}{r_o^3} \Bigg[-\frac{r_o^3}{3} + \frac{(ar_o^2+b)r_o}{a} -\frac{(ar_o^2+b)^2}{a^{3/2}}\tanh^{-1} \left(r_o \sqrt{a/(ar_o^2+b)}\right) \Bigg],
\end{split}
\label{eq:K}
\end{equation}

Back to Eq.~\eqref{eq:c_cr_sph}, we plug in $\beta$ and the exact solution of the Cr concentration profile becomes:

\begin{equation}
\begin{split}
    c_{\text{Cr}}(r) =& \alpha \ln \left( \frac{G}{6D_v} (r_o^2 - r^2)
    + c_v^{\text{eq}} \right) 
    + c_{\text{Cr}}^{\text{nominal}} -\alpha \ln (c_v^{\text{eq}})
     -\frac{ \alpha}{r_o^3} \Bigg[-\frac{r_o^3}{3} + \frac{\left(\frac{G}{6D_v}r_o^2 +  c_v^{\text{eq}}\right)r_o}{a} -\\
    &\frac{\left(\frac{G}{6D_v} r_o^2 +  c_v^{\text{eq}}\right)^2}{\left(\frac{G}{6D_v}\right)^{3/2}}\tanh^{-1} \left(r_o \sqrt{\left(\frac{G}{6D_v}\right)\bigg/\left(\frac{G}{6D_v}r_o^2 +  c_v^{\text{eq}}\right)}\right) \Bigg],
\end{split}
\label{eq:c_cr_sph_f}
\end{equation}

\section{Cr segregation profile in Spherical coordinates}
We compute the total Cr segregation toward the sink as:
\begin{equation}
\begin{split}
   c_{\mathrm{Cr}}(r) &= \alpha \ln\!\Big( A(r_o^2 - r^2) + c_v^{\mathrm{eq}} \Big) 
    + c_{\mathrm{cr}}^{\mathrm{nominal}} \\
    &\quad -\alpha \ln\!(c_v^{\mathrm{eq}})
     -\frac{\alpha}{r_o^3}\Bigg[-\frac{r_o^3}{3} + \frac{(A r_o^2 +  c_v^{\mathrm{eq}})r_o}{a} \\
    &\qquad -\frac{(A r_o^2 +  c_v^{\mathrm{eq}})^2}{A^{3/2}}\,
    \tanh^{-1}\!\Bigg(r_o\sqrt{\frac{A}{A r_o^2 +  c_v^{\mathrm{eq}}}}\Bigg) \Bigg],
\end{split}
\label{eq:c_cr_sph_f}
\end{equation}

\begin{equation}
\begin{aligned}
S^{Cr}_{sp} &\approx \int_{V} \big[c_{Cr}(r) - c_{Cr}(0) \big] dV \\
&
= 4\pi\int_{0}^{r_o} \big[c_{Cr}(r) - c_{Cr}(0) \big]  r^2 dr \\
&
=4\pi\alpha \int_{0}^{r_o} \bigg[ \ln \left( A (r_o^2 - r^2) + c_v^{\text{eq}} 
\right)-  \ln \left(B 
\right) \bigg] r^2 dr,
\end{aligned}
\label{eq:segregation}
\end{equation}

where for compactness we define
\[
A \equiv \frac{G}{6D_v}.
\]

\begin{equation}
B \equiv A r_o^2 + c_v^{\mathrm{eq}}, \qquad k \equiv \frac{A}{B}.
\end{equation}
Then
\[
A(r_o^2-r^2) + c_v^{\mathrm{eq}} = B - A r^2 = B(1 - k r^2),
\]
and the integrand simplifies:
\[
\ln\!\big(A(r_o^2-r^2)+c_v^{\mathrm{eq}}\big) - \ln B = \ln\!\big(1 - k r^2\big).
\]
Assume \(k>0\) and \(1-kr_o^2 = \dfrac{c_v^{\mathrm{eq}}}{B}>0\) so that the arguments of the log and \(\operatorname{atanh}\) are in the valid range.
Thus
\begin{equation}
\begin{split}
S^{Cr}_{sp} &= 4\pi\alpha \int_0^{r_o} r^2 \ln\!\big(1 - k r^2\big)\,dr\\
&=\frac{r^3}{3}\ln (1-kr^2) + \frac{2k}{3} \int \frac{r^4}{1-kr^2} dr\\
&=\frac{r^3}{3}\ln (1-kr^2) + \frac{2k}{3} \int \left[ -\frac{1}{k^2} -\frac{r^2}{k} + \frac{1}{k^2(1-kr^2)} \right] dr\\
&= \frac{r^3}{3}\ln (1-kr^2) - \frac{2r}{3k} - \frac{2r^3}{9k} + \frac{2}{3k^{3/2}} \mathrm{atanh} (\sqrt{k}r) + C.
\end{split}
\label{eq:S_simplified}
\end{equation}
Evaluate the definite integral from \(0\) to \(r_o\). At \(r=0\) every term vanishes, so
\begin{equation}
\begin{aligned}
I &\equiv \int_0^{r_o} r^2\ln(1-kr^2)\,dr \\
&= \frac{r_o^3}{3}\ln(1-kr_o^2)
- \frac{2r_o^3}{9}
- \frac{2r_o}{3k}
+ \frac{2}{3k^{3/2}}\,\operatorname{atanh}(\sqrt{k}\,r_o).
\end{aligned}
\label{eq:I_eval}
\end{equation}

\begin{align}
S^{Cr}_{sp}
&= 4\pi\alpha\Bigg[
\frac{r_o^3}{3}\ln\!\Big(1-k r_o^2\Big)
- \frac{2r_o^3}{9}
- \frac{2r_o}{3k}
+ \frac{2}{3k^{3/2}}\,\operatorname{atanh}\!\big(\sqrt{k}\,r_o\big)
\Bigg] \label{eq:S_final_a} \\
&= 4\pi\alpha\Bigg[
\frac{r_o^3}{3}\ln\!\Big(\frac{c_v^{\mathrm{eq}}}{B}\Big)
- \frac{2r_o^3}{9}
- \frac{2r_o}{3k}
+ \frac{2}{3k^{3/2}}\,\operatorname{atanh}\!\big(\sqrt{k}\,r_o\big)
\Bigg],
\label{eq:S_final_b}
\end{align}
where \(A=\dfrac{G}{6D_v}\), \(B=A r_o^2 + c_v^{\mathrm{eq}}\), and \(k=A/B\).

An equivalent form eliminating \(k\) is obtained by using \(1/k = B/A\) and \(1/k^{3/2} = B^{3/2}/A^{3/2}\):
\begin{equation}
\begin{split}
S^{Cr}_{sp}
&=4\pi\alpha\Bigg[
\frac{r_o^3}{3}\ln\!\Big(\frac{c_v^{\mathrm{eq}}}{\dfrac{G}{6D_v} r_o^2 + c_v^{\mathrm{eq}}}\Big)
- \frac{2r_o^3}{9}
- \frac{2r_o (\dfrac{G}{6D_v} r_o^2 + c_v^{\mathrm{eq}})}{3\dfrac{G}{6D_v}}\\
&+ \frac{2(\dfrac{G}{6D_v} r_o^2 + c_v^{\mathrm{eq}})^{3/2}}{3(\dfrac{G}{6D_v})^{3/2}}\,\operatorname{atanh}\!\Big(r_o\sqrt{\frac{\dfrac{G}{6D_v}}{\dfrac{G}{6D_v} r_o^2 + c_v^{\mathrm{eq}}}}\Big)
\Bigg].
\end{split}
\label{eq:S_final_c}
\end{equation}

\clearpage
\bibliographystyle{elsarticle-harv}
\bibliography{ref}